# Towards atomistic understanding of Iron phosphate glass: a first-principles based DFT modeling and study of its physical properties

Shakti Singh[a,*,‡], Manan Dholakia[a] and Sharat Chandra[a]

[a]Materials Science Group, Indira Gandhi Centre for Atomic Research, a CI of Homi Bhabha National Institute (Mumbai), Kalpakkam, TN, 603102 India

[‡]Present Address: Laser Biomedical Applications Division, Raja Ramanna Centre for Advanced Technology, a CI of Homi Bhabha National Institute (Mumbai), Indore, MP, 452013, India

E-mail: *corresponding author S.S: shaktisinghstephen@gmail.com,

M.D: manan@igcar.gov.on, S.C: sharat.c@gmail.com

## Abstract

Iron phosphate glasses (IPG) have been proposed as futuristic glass material for nuclear waste immobilization, anode material for lithium batteries and also as bioactive glass. In the last decade, there have been attempts to propose atomistic models of IPG to explain their properties from atomistic viewpoint and to predict their behavior in radioactive environment. In this paper, we seek to produce small scale models of IPG that can be handled within the scheme of Density Functional Theory (DFT) to study the electronic structure of this material. The starting models generated using Monte Carlo (MC) method [*S. Singh and S. Chandra, Comp. Mat. Sci.,* **202**, *110943, (2022)*] were subsequently annealed (at 1000 K) using ab-initio molecular dynamics (AIMD). This removes coordination defects present in the MC models. The equilibrated structure at this temperature was then force-relaxed using conjugate-gradient (CG) optimization. This hybrid approach (MC + AIMD + 0K DFT-CG optimization) produced good atomistic models of IPG which can reproduce experimentally observed electronic band-gap, vibrational density of states (VDOS), magnetic moment of Fe, the elastic constants as well as optical and dielectric properties. Computationally expensive melt-quench simulation can be avoided using present approach allowing the use of DFT for accurate calculations of properties of complex glass like IPG.

## I. Introduction

Iron phosphate glass (IPG) has attracted lot of research interest from glass scientists due to its superior properties that projects it as a next generation material for vitrification of high level radioactive waste from fast reactors [1–8]. Also recently its application as anode material in batteries was also investigated [9]. Iron phosphate glass optical fibers have also shown good mechanical properties [10] and recently iron containing phosphate glass fibers are also studied for slow-release fertilizers [11]. For understanding its properties in various settings, atomistic models that can reproduce already observed experimental results are vital. These models can then be further utilized to access properties that are tough to obtain experimentally through simulations such as ab-initio molecular dynamics (AIMD), classical molecular dynamics (CMD) and Monte Carlo methods. Overall, the last decade witnessed a few simulation [12–17] as well as experimental studies (recent [18–22] and older [23–30]) to explore IPG as waste storage material.

In this paper, we further the research on this material by designing atomistic models with periodic boundary conditions (PBC) that can be utilized in first-principle based codes due to their smaller size. First-principle based methods have not been explored "properly" for this material for demand of large computational resources and time required in model production through melt-quench simulation technique. In the present study, we report a new method to model complex glassy materials which relies on Monte-Carlo method [31] to produce random-network starting structure which is then annealed at high temperature to remove most of the coordination defects and subsequently thoroughly force-relaxed (for ionic forces and overall external pressure on the simulation cell) at 0 K by conjugate gradient relaxation. This promises an overall strain-minimized model that has been shown to reproduce most of the experimentally known properties.

Initial studies on IPG focused on finding the structural connections involved in the random network of constituents. Al-Hasni and Mountjoy [32] used MD simulation to model IPG employing the melt-quench (MQ) way to reach the amorphous structure. This was followed by Stoch *et al.* [20,33] who proposed a cluster based model, followed by density functional (DFT) study of various structural connections found in crystals of similar compositions. Then, using the interatomic potentials developed by Jolley and Smith [14], Kitheri *et al.* [16,17] came up with larger scale models of IPG obtained from melt-quench simulation in classical molecular dynamics and also carried out experiments to determine many properties of IPG [21,22]. Even though large-scale models developed using CMD are necessary for some complex simulations like radiation damage, when it comes to obtaining the electronic properties accurately, models must be prepared using ab-initio approaches. In order to estimate properties accurately, in the present work, we used DFT to study the electronic structure of IPG as well as its mechanical and optical properties. Further the present study explores different exchange-correlation functionals and the resulting change in its structure and properties. These models can also be used to refine inter-atomic potentials used in CMD simulations, for simulating processes at large length and time scales, such as radiation damage in radioactive waste storage materials.

In this study, modeling from first principle allows for accurate estimation of the electronic structure as well as other properties which were not reported till now for this material. Second, circumventing MQ simulation (meant for randomizing the starting structure and quenching the disordered state) results in making the modeling process computationally light comparatively. Third, the in-house developed potential-free MC method used for producing randomized initial configuration of glass is customizable and can be used to explore models with different compositions of FeO, Fe$_2$O$_3$ and P$_2$O$_5$. At present such studies are dependent on either using only crystalline systems or using totally random atomic positions as starting point in melt-quench simulation.

The paper is organized as follows: Section II presents the hybrid approach followed in this study to arrive at IPG glass models and the details about the models developed in this study are also presented. Section III describes the results and discusses in (1) the structural properties viz. radial distribution function (RDF), coordination numbers, bond angles, static structure factor, rings distribution and void distribution of the models. In Section III (2) the electronic structure of IPG is presented. Section III (3) outlines the mechanical properties of IPG i.e. the vibrational density of states and elastic constants, section III (4) presents the optical properties, followed by the conclusions in section IV.

## II.   Model generation: hybrid method: MC + high T annealing at 1000 K in AIMD + CG relaxation at 0 K

Developing models of glasses is a non-trivial problem in glass research and although the melt-quench simulation is most widely used nowadays, issues like unphysical quench rate, lack of initial glass configurations and computational bottlenecks associated with ab-initio approaches have limited the research. In this study, a hybrid approach is used to model IPG as shown in steps 0-3 in Figure *1*. Here step 0 shows the randomized PBC atomistic model produced from potential-free MC method [31] containing 201 atoms (136 O, 39 P, 26 P) in a cubic simulation cell of edge 14 Å (dictated by its experimental density) as initial configuration. It is then annealed at 500 K and 1000 K for 5000 femtoseconds in NVT ensemble using AIMD (using DFT-VASP [34,35]) to anneal out most of the coordination defects as shown in step 1. Subsequently in step 2a, to reach the ground state corresponding to this annealed model, the model is relaxed using CG optimization so that the forces on individual ions reach to below 0.01 eV/Å. For reaching correct density corresponding to the ground state and to minimize the external pressure on the simulation box, Birch-Murnaghan (BM) equation of state [36] fitting is used to obtain the optimal lattice parameter, followed by a final step of ionic relaxation at the optimal lattice constant. This further promises a minimum strained structure and completes the step 2b of calculations. The energy functional used in the DFT calculations outlined above is simple PBE functional and until mentioned otherwise this is the go-to functional for most calculations. Finally, the optimized structure corresponding to different density functionals i.e PBE, PBE + U, HSE and HSE + U are then obtained by only relaxing the ionic degrees of freedom. The value of cut-off energy used for limiting the number of plane waves is 500 eV and only gamma (Γ) point simulation is done due to large simulation cell with no symmetry. For the case of simulation done using the U parameter, it is set at 3.5 eV for the Fe atom. For the case of HSE06 calculations, the default values of mixing parameters are used. This corresponds to using 25% of the exact exchange energy from Hartree-Fock-type calculations, 75% from semilocal-exchange from PBE calculations and 0.2 Å$^{-1}$ as the screening length. The electronic density of states (EDOS) plots are obtained on an increased K-mesh sampling of 4x4x4 and its convergence with respect to K-sampling is also tested. Spin polarized calculations are performed for studying the magnetic behavior. The ionic step stopping criterion for geometry relaxation calculations is when the forces become less than 0.01 eV/Å for each atom inside the simulation cell.

In reporting all the calculated properties, care has been taken to compare them with experimental data wherever available, so that these models get validated against the vast experimental data already reported for IPG since the last two decades.

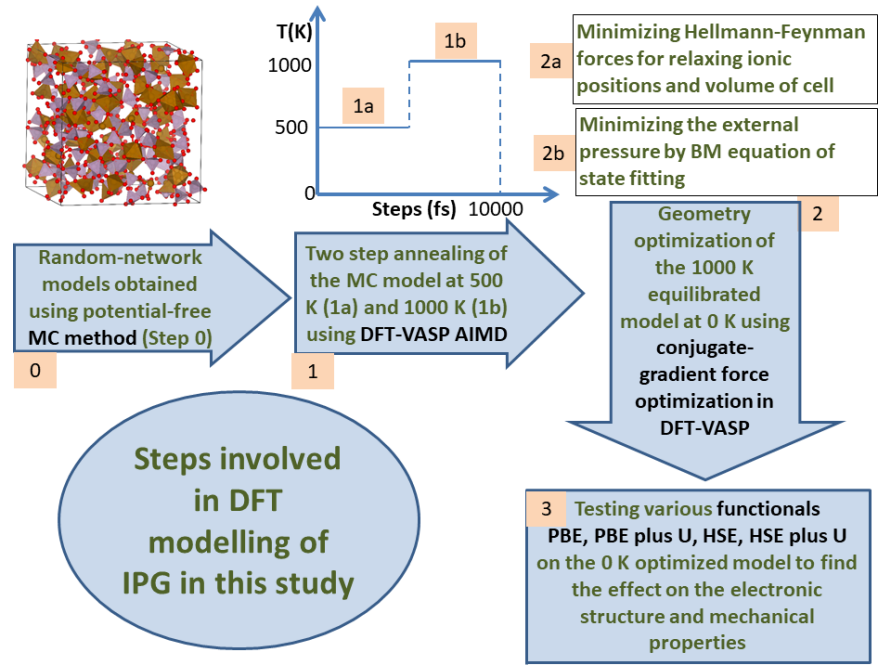

Figure 1: Pictorial schematic: Thermal treatment of MC models and final CG relaxation and BM equilibrium density estimation at 0K

### III. Results and Discussion
### 1. Structural Properties

### A. RDF analysis

RDF is the simplest yet most effective tool for studying amorphous/glass networks. Bond lengths can easily be determined from inspection of RDF and partial pair correlation functions. For small-scale models such as those developed in this study, the RDF curves do have fluctuations/noise. In large-scale models, the fluctuations/noise average out and give smooth curve due to the largeness of sample space of possible bond-distances. In small-scale models, sufficient accuracy in determining the bond-lengths can also be achieved by smoothening the curve. In this way, one can estimate the true mean from sample mean. The small sample may lead to larger standard deviations around the mean but still the mean bond-lengths will give a good estimate of the realistic bond-lengths. RINGS software [37] is used to calculate the radial distribution function g(r) and other structural properties reported in this study. Experimentally the quantity that is reported for glasses is total pair correlation function T(r), so in Figure *2* T(r) of various models are compared with experimental T(r) obtained in [23]. The relation between various definitions of pair correlation function reported here is provided in supplementary file.

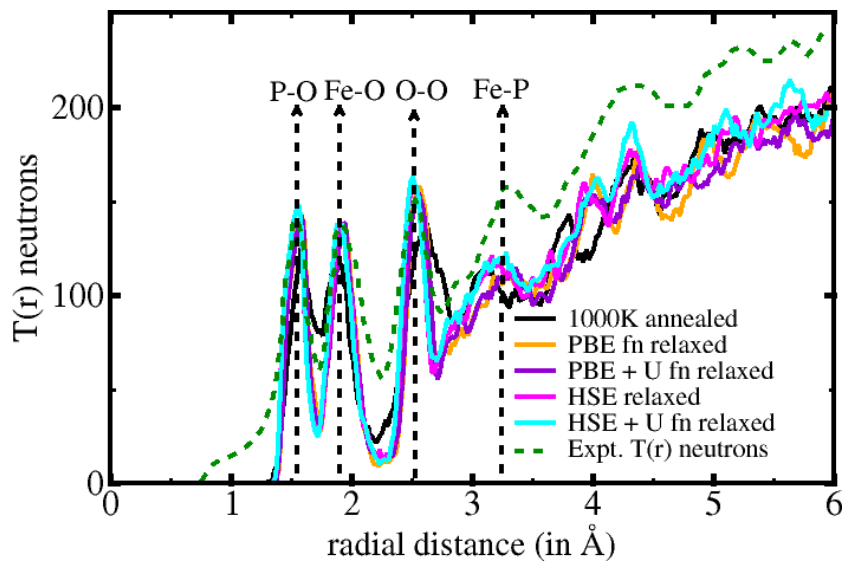

Figure 2: The total neutron pair correlation function T(r) for various stages of the developed IPG models namely 1000 K annealed (black), PBE XC fn relaxed (orange), PBE + U fn relaxed (violet), HSE relaxed (magenta), HSE + U fn relaxed (cyan), and the experimental T(r) (dotted green)[23]

In Figure 2, the comparison of total neutron pair correlation function T(r) for various stages of the developed IPG models is shown. In particular, the models considered for comparison are: the model obtained after annealing the sample at 1000 K in AIMD labeled as "1000 K annealed model", model obtained after relaxing ionic position and volume of the 1000 K model with PBE exchange-correlation function and conjugate-gradient optimization scheme at 0 K. This model is labeled "PBE functional relaxed model". Next, the PBE functional relaxed model is further relaxed by accommodating the U correction in the XC functional, labeled as "PBE + U functional relaxed model". Similarly, the model relaxed using hybrid exchange-correlation functional HSE06 is labeled "HSE relaxed model" and that with U correction as "HSE+U relaxed model". Finally the experimental value of T(r) is also plotted from [23]. We can see that the T(r) peak positions produced by our structure matches well with that of the experimental value. The first four prominent peaks can be identified with P-O (1.51Å), Fe-O (1.8 Å and 1.94 Å), O-O (2.48 Å and 2.74 Å) and Fe-P (3.3 Å) bond distances, also labeled with vertical dotted lines in Figure 2.

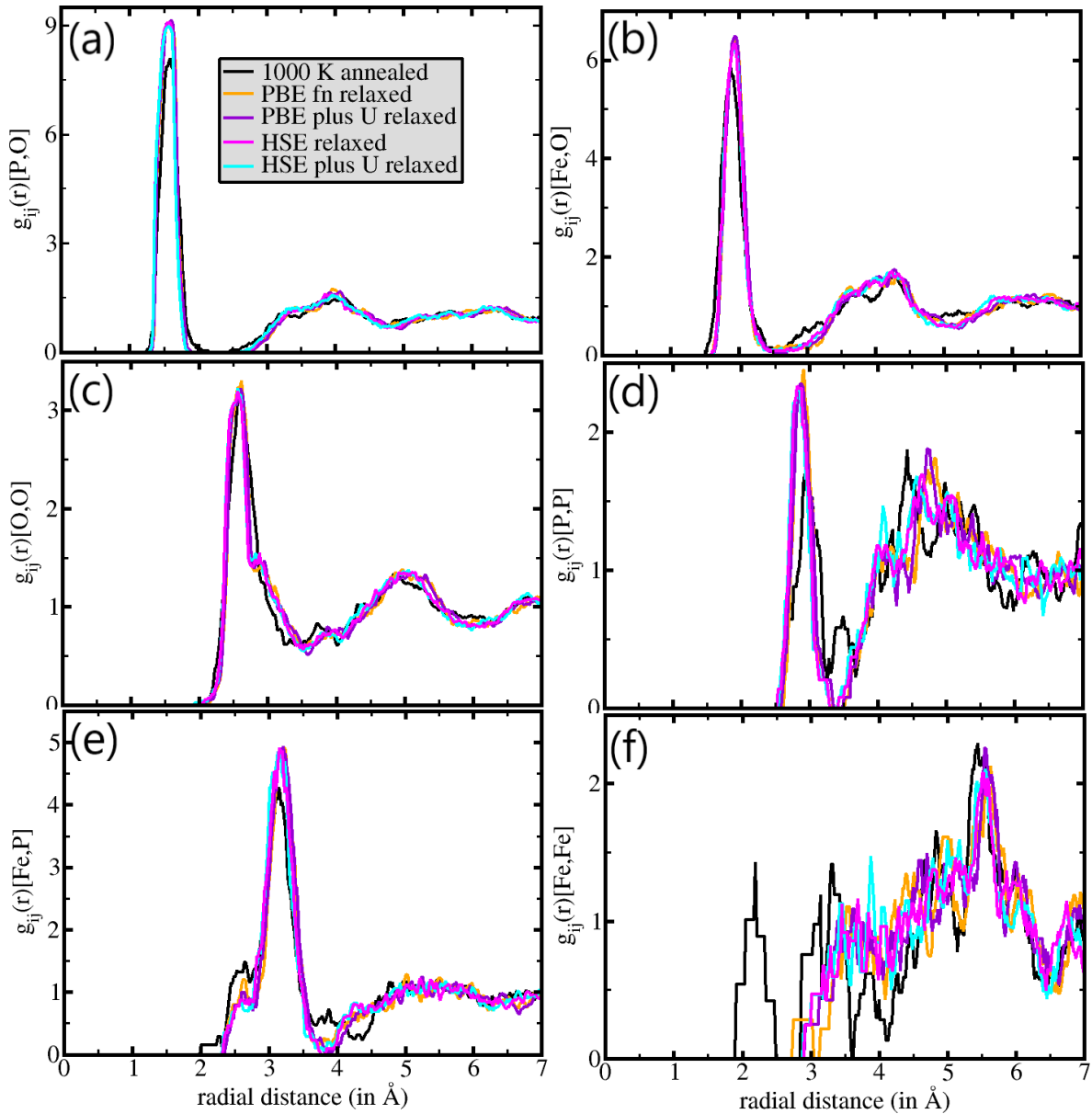

Figure 3: Graphs showing the comparison of partial pair distribution function of the developed models

Figure 3 shows a similar comparison among the models with respect to partial pair correlation functions. Six possible pair correlation functions in IPG are shown. They give a clearer picture to the peaks discussed in total correlation function T(r). The partial pair distribution functions bring out more prominent correlations in the models, which gets suppressed in T(r) due to fluctuations. In (a) $g_{ij}(r)[P,O]$, the peak at 1.51 Å corresponds to P-O bond distance. In (b) $g_{ij}(r)[Fe,O]$, the peak at ~1.9 Å engulfs two bond distances within it i.e. Fe-O bond distance of 1.8 in tetrahedral coordination geometry and 1.94 in octahedral coordination geometry. The upcoming bond distances then form the next nearest neighbor distances, and the corresponding peaks are broader and less intense. In (c) $g_{ij}(r)[O,O]$, the peak at ~2.5 Å corresponds to O-O distance in tetrahedral vertices, whereas the subdued peak or rather shoulder at 2.75 corresponds to O-O distance between oxygens at octahedral vertices. In (d) $g_{ij}(r)[P, P]$, the peak at ~3.0 Å is from P-P correlations. In (e) $g_{ij}(r)[Fe,P]$, the peak at ~3.27 Å is from Fe-P correlation between the Fe octahedral and P tetrahedral units. One can also observe a small peak coming from the edge-shared Fe-P correlation appearing at ~2.7 Å. Lastly the Fe-Fe correlation builds around 5.5 Å in (f) $g_{ij}(r)[Fe, Fe]$ with high fluctuations due to higher order of these correlations.

These values can be compared with the experimental values of these correlations from [23]. The values are: P-O = 1.53 Å, Fe-O = 1.94 Å, O-O = 2.52 Å, Fe-P = 3.23 Å. The close agreement of values obtained from simulation validates the structural parameters of the obtained models.

### B.  Coordination number (CN) analysis

There are three coordination geometries to be considered here: coordination of network former P, network former/modifier Fe and the bridging element of network O.

From the literature [30,38], we have information that Fe is present in two oxidation states i.e., 3+ and 2+. $Fe^{3+}$ is present in both tetrahedral and octahedral environments, with $Fe^{2+}$ favoring a distorted octahedral geometry. Fe acts as network former when it forms tetrahedron geometry with a CN of 4, whereas it takes an octahedral or distorted octahedral or square pyramidal geometry while acting as network modifier with a CN of 6 or 5. This intermediate nature of Fe [39] which is mainly due to the ionic nature of the bonding, leads to its average CN lying between 4 and 6. Phosphorous on the other hand, is the main network former in this glass, and prefers to occupy tetrahedral geometry only due to covalent nature of bonding (CN = 4). The bridging element oxygen has an average CN of 2 in a perfect network but can deviate from this value due to coordination defects such as edge-sharing, face-sharing or dangling bond.

Table 1 and Table 2 provide the distribution of various coordination geometries of P and Fe, respectively in different models of IPG considered in this study. The values are compared with experimental data and MC-MD model obtained from classical MD equilibration simulations on Monte-Carlo model as described in [31,40]. The agreement between the calculated and experimental value validates our models obtained in this study. One very interesting observation from the CN analysis is that while the P-O tetrahedra is very rigid due mainly to its covalent nature (as seen from very small percentages of coordination geometries of P with CN 3 and 5), the Fe-O coordination geometries are flexible due to its ionic character (as seen from high percentage of geometries with CN 5). This high percentage of Fe in CN 5 geometry is also in contrast to what is generally seen in iron-phosphate crystals that IPG mostly crystallizes into i.e $Fe_3(P_2O_7)_2$ and $Fe_4(P_2O_7)_3$. This is hence a characteristic of IPG. In supplementary file, some snapshots of the model showing different coordination geometries is given in Figure S1. The distribution of Fe in different CN states differs from classical MD results, in which the concentration of tetrahedral geometry is found to be 61% followed by 5-coordinated Fe at 31% and very small 7% of Fe in octahedral geometry. We also did the CN analysis of a huge 90 Å cubic model of IPG prepared by melt-quench simulations in MD [13]. The distribution is shown in last column of Table *2*. The agreement among various models discussed in this study with classical MD results is quite good for P, but for Fe the deviation of percentage of units in various coordination in models developed using MD as well as the difference in average CN of Fe warrants further research into the interatomic potentials of IPG used in MD, which is outside the scope of the present study. Figure 4 shows the variation of average coordination numbers of Fe, O and P across different models of IPG in a bar chart.

Table 1: Phosphorous coordination geometry and its distribution in different models of IPG considered in this study

| CN of P | % distribution of such configurations | | | | | | | |
|---|---|---|---|---|---|---|---|---|
| | MC model | 1000 K model | PBE relaxed model at 0 K | PBE+U relaxed model | HSE relaxed model | HSE +U relaxed model | MC-MD equilibrated model [40] | MD-MQ model [13] |
| 0 | - | - | - | - | - | - | - | - |
| 1 | - | - | - | - | - | - | - | 0.36 |
| 2 | 2.56 | - | - | - | - | - | - | 2.27 |
| 3 | 2.56 | 7.69 | 2.56 | 2.56 | 2.56 | 2.56 | - | 5.26 |
| 4 | 94.87 | 92.31 | 94.87 | 94.87 | 94.87 | 94.87 | 93.16 | 92.11 |
| 5 | - | - | 2.56 | 2.56 | 2.56 | 2.56 | 6.84 | - |
| **Average behavior** | | | | | | | | |
| Avg. CN | **3.92** | **3.92** | **4.0** | **4.0** | **4.0** | **4.0** | **4.07** | **3.89** |
| Expt. | **3.8 ± 0.2 (XRD [23]), 3.6 ± 0.2 (ND [23])** | | | | | | | |

Table 2: Iron coordination geometry and its distribution in different models of IPG considered in this study

| CN of Fe | % distribution of such configurations | | | | | | | |
|---|---|---|---|---|---|---|---|---|
| | MC model | 1000 K model | PBE relaxed model at 0 K | PBE+U relaxed model | HSE relaxed model | HSE +U relaxed model | MC-MD equilibrated model [40] | MD-MQ model [13] |
| 0 | - | - | - | - | - | - | - | - |
| 1 | 3.85 | - | - | - | - | - | - | 0.21 |
| 2 | 30.77 | - | - | - | - | - | - | 1.67 |
| 3 | 34.62 | 7.69 | - | - | - | - | 1.39 | 5.70 |
| 4 | 19.23 | 11.54 | 34.62 | 30.77 | 30.77 | 26.92 | 61.11 | 46.73 |
| 5 | 3.85 | 61.54 | 53.85 | 53.85 | 50.00 | 50.00 | 30.56 | 41.15 |
| 6 | 7.69 | 19.23 | 11.54 | 15.38 | 19.23 | 23.08 | 6.94 | 4.52 |

| Average behavior | | | | | | | | |
|---|---|---|---|---|---|---|---|---|
| Avg CN | 3.12 | 4.92 | 4.77 | 4.85 | 4.89 | 4.96 | 4.43 | 4.41 |
| Expt. | 4.7 ± 0.3 (XRD [23]), 4.6 ± 0.3 (ND [23]), 4.8 (EXAFS [27]), 4.74 ± 0.14 (EXAFS [41]) | | | | | | | |

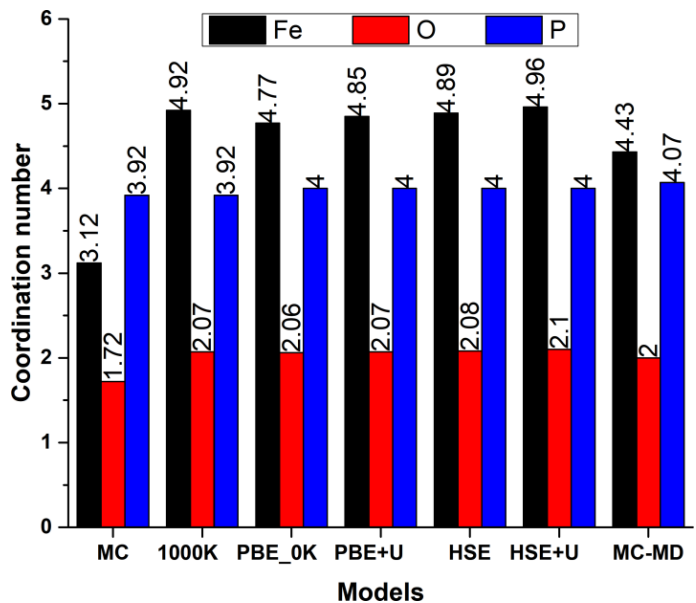

Figure 4: Coordination number of Fe (black), O (red) and P (blue) and its comparison for different models

### C. Bond-angle distributions

Figure 5 shows the comparison of 6 possible bond angles in IPG structure. The O-Fe-O bond angle distribution (BAD) in our MC structure shows three distinct peaks around 90°, 109.5° and 180°. In an ideal Fe octahedral geometry (CN = 6), there are 12 O-Fe-O angles equal to 90° and three O-Fe-O angle as 180°. Due to this, there is large difference in the bond-angle percentages around 90° and 180°. The presence of tetrahedral motif (CN = 4) results in peak around 109.5° which gets merged with 90° peak, when the values are fitted with a polynomial function. Also, significant percentage of Fe are in a distorted coordination geometry (CN = 5). Analyzing such geometries in our structure, it was observed that Fe is in either distorted trigonal biprism geometry or distorted square pyramidal geometry. These geometries also give three possible distributions of O-Fe-O bond angles, namely around mean values of 90°, 120° and 180°. Overall, when fitted with a polynomial function, the BAD shows a main peak at 97° due to larger percentage of angles around this value (namely 90°, 120° and 109.5° from the above discussion) and a smaller peak at 160° (from angles around 180°).

The O-P-O BAD peaks at 109.5°, as expected for the $PO_4$ tetrahedral motif. All the models have peaks around this mean value, which compares well with the model obtained from MC-MD hybrid approach. P-O bond being covalent in nature resists any significant deviation from its general tetrahedral coordination geometry.

The Fe-O-Fe BAD peaks are observed in range 130°-140° for various models in this study. The value of peak for MC-MD model is at 112.9°. This brings out a crucial difference between bond-angles obtained from ab-initio studies from that of bond angles obtained from MD studies that are dependent on empirical inter-atomic potentials. Also, there is a clear difference of 10° in BAD peaks obtained with and without U correction. This also brings out the important point that U correction is very important for correct orientation of Fe atoms w.r.t each other. This will lead to more accurate prediction of magnetization with U correction in the functional.

Similarly, there is a difference in P-O-P BAD. The peak of this distribution is at ~123° for our structure, whereas the MC-MD model gives a peak at ~149.8°. The mean value of this angle distribution obtained in this study agrees closely with value of P-O-P bond-angle (125.5°) obtained in a study done on iron-phosphate clusters using DFT [20].

Finally, the P-O-Fe BAD for our structure shows mainly two peaks, the main peak corresponding to 135° and small peak at 80° seen as shoulder in the fitted curve. The 80° peak may be due to the presence of small percentage of Fe in 2+ oxidation state which sit in sterically hindered position between two $Fe^{3+}$ atoms. The 135° peak comes from the usual connection between Fe octahedral or tetrahedral motif with P tetrahedral motif. The MC-MD model has a peak at 138.7° which is in good agreement with the BADs of ab-initio obtained models. Overall, the bond-angles are in good agreement with MC-MD model, with the only significant deviation seen in P-O-P BAD, the reason for which is not yet well understood.

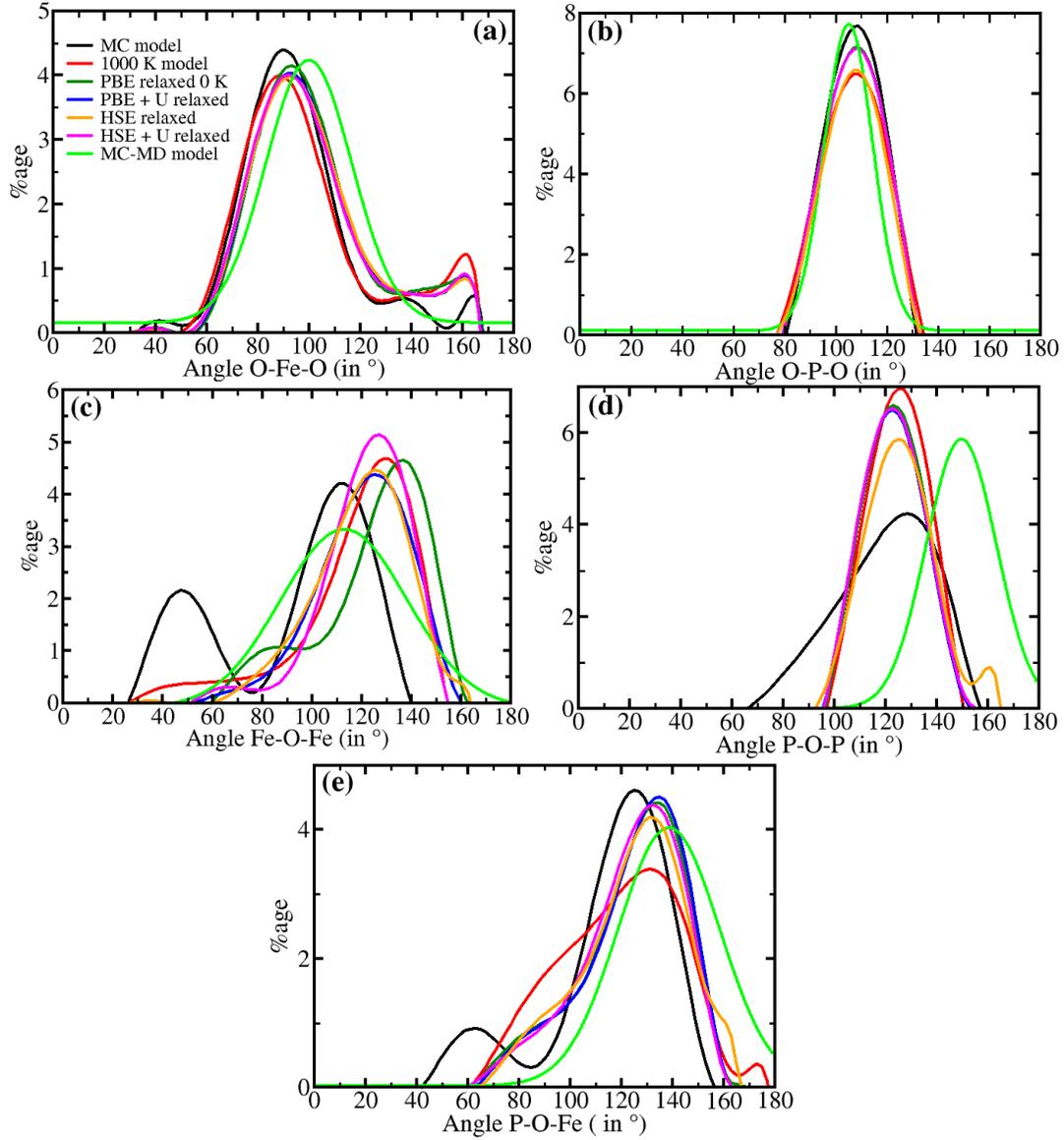

Figure 5: Graphs showing angle distributions in the structure of IPG developed in this study as compared with MC-MD generated structure from [31]

## D. Static structure factor analysis and FSDP

The calculated structure factor of IPG gives a good quantitative as well as qualitative idea about the disorder in the developed models. In particular, the first peak often called first sharp diffraction peak (FSDP) which has been a universal feature of glasses is routinely linked to the presence of intermediate-range order (IRO) in glasses. Here the calculated neutron structure factor of our models is compared with existing theoretical models (MC-MD model from [31] and MD-MQ model from melt-quench simulations from [13]) and experimental data [23] in Figure 6. The models obtained in this study, though smaller in size as compared to the MD-MQ model, satisfactorily reproduce the FSDP at correct value of scattering vector. The discrepancy in terms of the intensity of FSDP i.e. the DFT produced models underestimating the maximum of FSDP while classical MD produced models overestimating the peak intensity, can be due to the arbitrary value reported experimentally in [23]. Upon visual inspection, it is further observed that the subsequent peaks in S(q) shown by dotted vertical lines at 3, 3.7, 4.25, 5 and 6.25 Å$^{-1}$ are better reproduced by the DFT models than the models produced using classical MD.

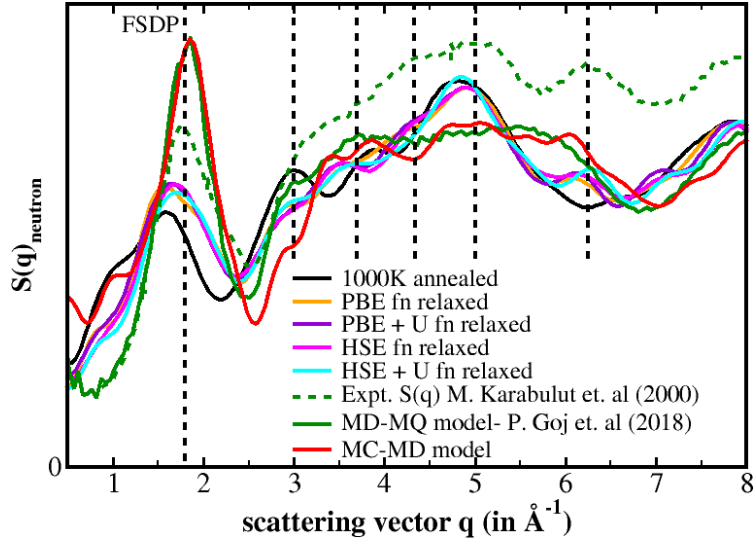

Figure 6: Calculated neutron structure factor for various models and comparison with S(q) from experiment [23] as well as for a model from MD melt-quench (MQ) simulation [13] and model obtained from just MD equilibration of MC model (MC-MD model [40])

### E. Rings distribution

Identifying the medium-range order in glasses has also been a central topic in glass research and recently the topological data analysis methods like finding the rings distribution has gained attraction [42,43]. The R.I.N.G.S. software [37] is used in this study to find the distribution of different ring sizes present in the atomistic models. The software provides a few possibilities of this calculation based on different definitions of rings. Here, we have reported Guttman's [44] rings distribution. In this definition, a ring is defined as the shortest path between two of the nearest neighbors of a given node atom.

Figure 7 shows the calculated rings distribution of the models discussed in this study. The rings distribution is found to skew towards smaller sized rings due mainly to the small size of these models which restricts the formation of large sized rings since the periodic boundary condition sets in at smaller length-scales. Since rings distribution is a theoretical construct and lacks any experimental observation against which it can be validated, it has been compared only with rings distribution of existing simulation model developed using hybrid approach combining MC+MD as explained in [40]. Although DFT is immensely useful for electronic structure analysis of systems, yet for amorphous systems, due to absence of long-range order, the choice of sufficiently large simulation cell is highly restricted. In order to simulate bulk, smaller cells when used in PBC, lead to inherent order which is absent in realistic amorphous systems. This is correctly brought up by the rings distribution analysis, which shows that the limitation of the presented models in this study to produce the rings distribution as seen in MD study of IPG glass. In the absence of any experimental probe that can be used to correlate to rings distribution, the correctness of this property is subject to scrutiny.

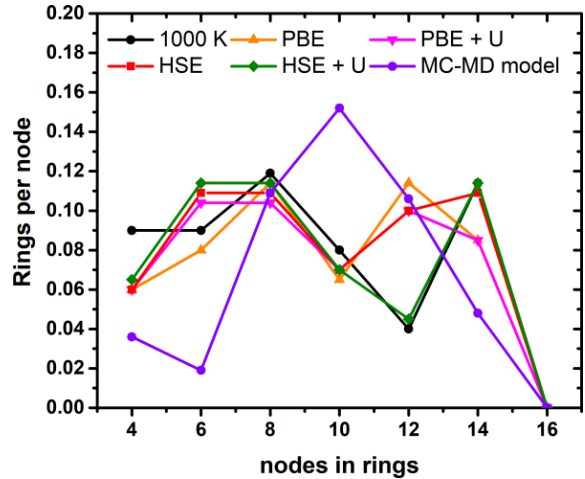

Figure 7: Rings distribution of IPG models calculated using RINGS code

### F. Void analysis

The void distribution of simulation models is an important tool which can be used to judge whether the developed model is homogeneous or not. Assuming the atoms in the model to be spherical with some radius (could be atomic, ionic or van der Wall's), the empty space in the model can be filled with spherical voids of varying radii. Generally, any proposed simulation model of glass should be devoid of any abrupt peak in void size distribution. The distribution should be like a Gaussian one, with low concentration of small as well as large sized voids and high concentration of medium-sized voids. This is unlike crystals where depending upon the lattice one can have multiple peaks associated with different size of the voids, for example we have two peaks in the void distribution of face-centered cubic system, corresponding to the tetrahedral and octahedral voids.

Table 3 provides the density and void fraction of different models considered in this study. Also provided are the values of these long range dependent properties for MD models from literature [13,31]. All the models considered in this study have density close to the experimental density of IPG which is reported to be $3.0 \pm 0.1$ g/cc (references given in Table 6). Figure 8 presents the void size distribution of the considered models, all conforming to the basic understanding of the size of voids inside glass discussed above. All the relaxed models show a different slope of the void distribution after a diameter of 2.5 Å. The knowledge of the size and position of these larger voids can be very important in making an informed choice about the type of radioactive element that can be placed inside a void for radiation-related studies on IPG.

Effect of functionals on void-fraction and voids: All the functionals used for relaxing the forces on glassy models result in decreasing the void fraction in the model. This can be understood in terms of improved packing of motifs upon structural relaxation of these models. On the other hand, the density is found to decrease for the PBE+U functional model while the HSE+U functional increases the density.

Table 3: Variation of density and void fraction for different functionals employed and a comparison with MC-MD model

| IPG Model | Density (g/cc) | Edge of cubic cell (Å) | Void fraction (dimensionless) |
|---|---|---|---|
| MC model | 2.93 | 14.0 | 0.299 |
| PBE | 2.80 | 14.21 | 0.284 |
| PBE + U | 2.79 | 14.24 | 0.287 |
| HSE | 2.93 | 14.0 | 0.265 |
| HSE + U | 2.99 | 13.9 | 0.253 |
| MC-MD model [31] | 2.91 | 20.0 | 0.313 |

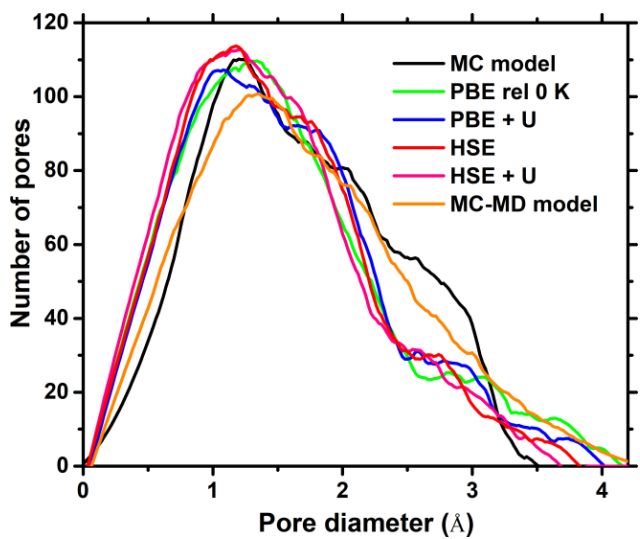

Figure 8: Void size distribution of IPG models

2. **Electronic structure**
A. **EDOS, IPR and Band-gap**

The EDOS, inverse participation ratio (IPR) and band-gap of IPG calculated using different density functionals are shown in Figure 9. The figure plots EDOS and IPR in stacked arrangement for (a) PBE, (b) PBE + U, (c) HSE06 and (d) HSE06 + U functionals. The band-gap across the Fermi level is also marked with dotted vertical lines done only through visual inspection (guided by fermi energy and band-tailing due to disorder). The estimation of band-gap purely from visual inspection of EDOS and IPR is difficult for amorphous/glassy systems due to the presence of defective coordination geometries that contribute to the electronic density in the gap region and also leads to tailing (exponential decay of electronic density) of band-edges [45,46]. The existence of these states has been shown for many amorphous systems using computational models and experiments [47–49]. A more accurate determination of the band-gap is also done through estimation from the Tauc's plot in section VI. Following observations from Figure 9 require attention:

(a)     The band-gap of IPG is dependent on the functional used for relaxation of the model. While PBE and PBE + U functional underestimate the band-gap, HSE and HSE + U functional overestimates it. Calculation with HSE functional produces the closest agreement with experimental band-gap (see Table 4 for comparison).
(b)     From the IPR plot, one can see the presence of localized electronic states (states having high IPR) near the band-tails, in the gap region as well as in the conduction band.
(c)     The U correction is important for calculations involving transition elements and results in more splitting of the valence and conduction band across the Fermi level. More importantly for the correct estimation of magnetism in IPG, the U correction is found to be very useful.
(d)     The states deep in the bands have a small IPR representing them as delocalized electronic states.
(e)     In Figure 9(d), two band-gaps are marked. The one using dotted black lines is that from visual inspection and the one using dotted green lines is the one obtained from the Tauc's plot in section VI. The close agreement between the two values also validates our markings done using visual inspection only.

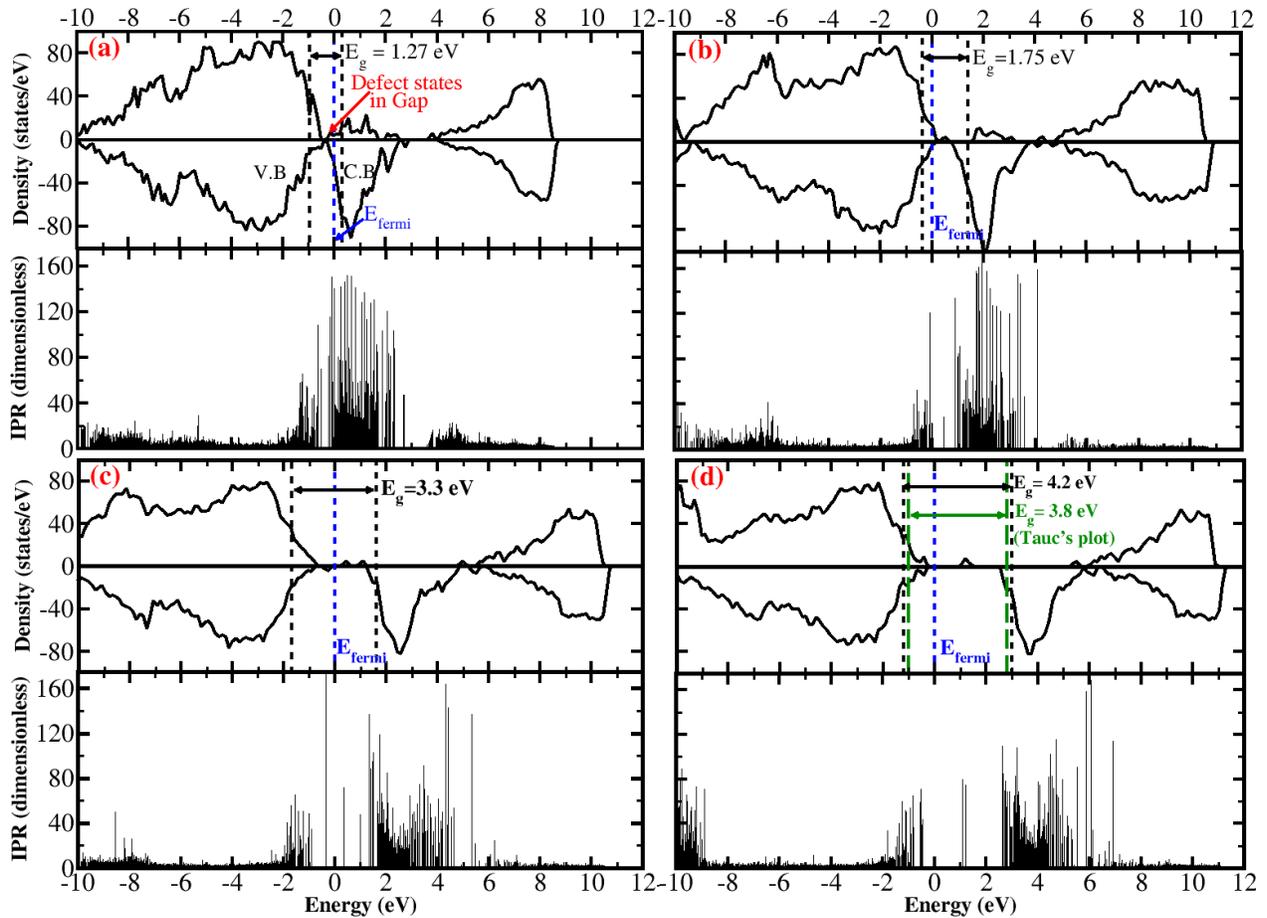

Figure 9: Total DOS (top) and inverse participation ratio (bottom) of (a) PBE functional relaxed model (b) PBE + U fn. relaxed model (c) HSE06 relaxed model and (d) HSE06 + U relaxed model of IPG

The partial DOS (PDOS) showing the contribution from different orbitals of Fe, P and O is plotted in Figure 10. Following observations can be made from the figure:

(a) O-*s* contributes mainly to the formation of deeper bands below valence band (VB). Whereas O-*p* orbital contributes mainly to the valence band.
(b) The P-*p* contributes to states in the conduction and valence bands but away from the CB and VB edges.
(c) The Fe-*d* contributes to states near the band-gap edges. This also highlights the importance of U correction in PBE based DFT calculations of IPG since the *d* electrons of Fe have high correlation energy. The accommodation of U correction is less effective in HSE calculations since the hybrid functionals are known to produce the exchange part exactly and correlation part approximately through mixing of the Hartree-Fock exchange and PBE functional. This can be seen from band-gap and magnetism estimation.

Hence, from the PDOS graph (Figure *10*) it is concluded that optical band-gap formation in IPG is determined mainly from the separation of the transition metal Fe *3d* states and the *2p* oxygen states, both of which contribute towards the valence and conduction band. This finding is also corroborated by experimental studies by ML Schmitt [50].

Table *4* lists some electronic properties of IPG obtained from different functionals and their comparison with the experimental data.

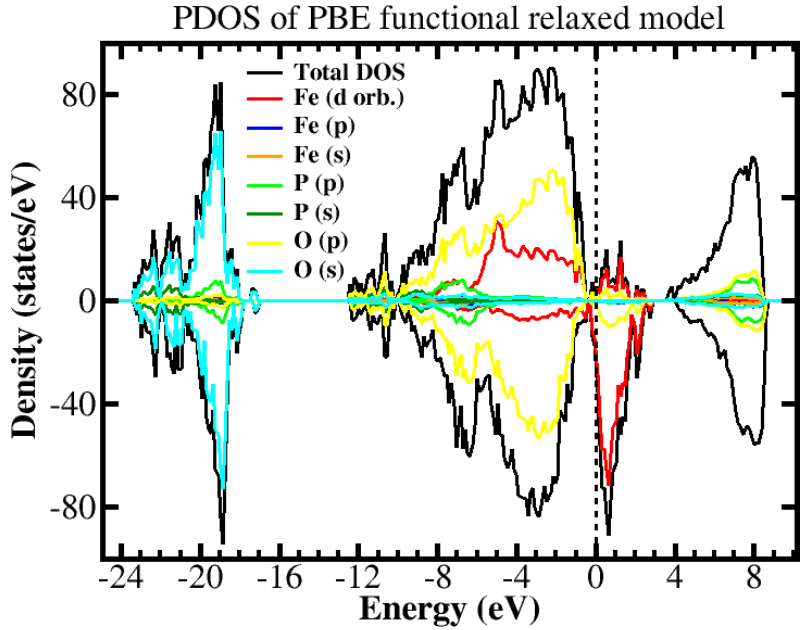

Figure 10: Total and atom-decomposed partial DOS of IPG obtained using standard PBE functional.

Table 4: Variation of electronic property with change in exchange-correlation functional used for relaxing the model and comparison with experimental data

| Property | PBE functional (standard DFT) | PBE + U corr. | Hybrid functional | Hybrid + U corr. | Experimental data |
|---|---|---|---|---|---|
| Total energy (eV/atom) | -7.22 | -7.05 | -9.54 | -9.41 | - |
| Average Magnetization on Fe ($\mu_B$) | 2.9 | 4.34 | 4.13 | 4.54 | 5.64 [51] |

| Electronic Band-gap, $E_g$ (eV) | 1.27 | 1.75 | 3.3 | 4.2 | 2.85 [9], 2.9 [52], 2.9-3.2 [50] |

## B. Magnetization

The average magnetization on Fe atom is found to be dependent on the functional chosen as well. Closest agreement with the experimental value of average magnetization of 5.64 $\mu_B$ [51] is obtained with HSE functional together with U correction as reported in Table 4. U parameter is found to have a crucial impact on the magnetization, as seen for calculations with U and without U correction in PBE functional relaxed model, whereas with HSE functional the impact of U correction is found to be minimal. Values of magnetic moments on individual atoms are reported in Supplementary file (Table S1 and S2). In Figure 11, the atoms with finite magnetic moment are shown which happen to be only Fe atoms. The length of vectors shown on these atoms are proportional to the magnetic moment estimated from DFT calculations. The PBE functional and PBE + U functional relaxed models show 3 and 2 Fe atoms, respectively in antiferromagnetic orientation (shown by green vectors in Figure 11(a) for the case of PBE relaxed model). Experimental studies [29,30,53] suggest that IPG does show short-range antiferromagnetic (speromagnetic) ordering at low temperature. Although such a behavior cannot be concluded from the present study, the presence of magnetic moments with varying magnitude and also in opposite direction, confirms presence of spin-glass like magnetism in IPG. The HSE functional relaxed model on the other hand shows only ferromagnetic ordering as shown in Figure 11(b).

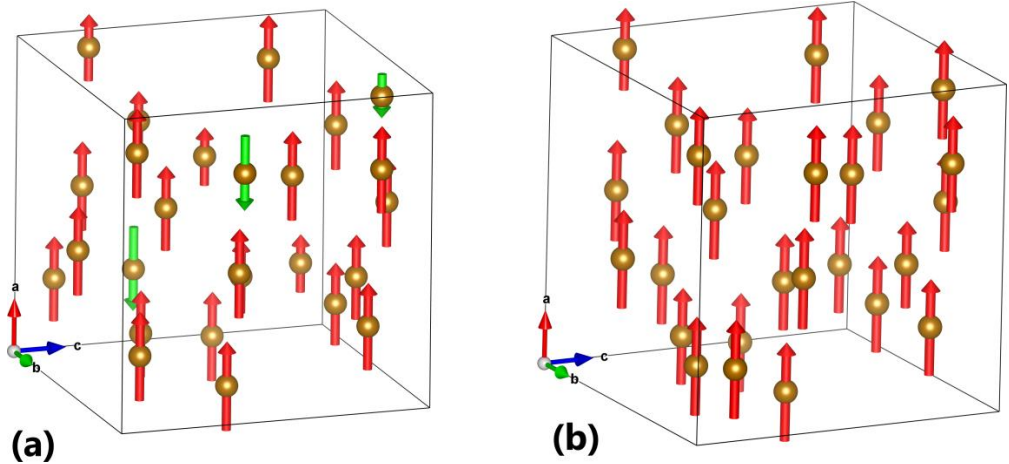

Figure 11: Magnetization vector on Fe atoms in the final IPG models obtained after (a) PBE functional relaxation and (b) HSE functional relaxation

## 3. Mechanical properties

## A. Vibrational DOS

The vibrational DOS (VDOS) of our PBE relaxed model is presented in Figure 12. The VDOS is calculated using Phonopy code [54] from DFT calculations. Using the finite difference method, second order force constants are determined when each ion is displaced in each independent direction. This is called "frozen phonon approach" to calculate the zone-center vibrational frequencies of a system. It requires calculating the total energies for at least two displacements in the $\pm x$, $\pm y$ and $\pm z$ directions for each atom as the system has only the translational symmetry. For the present system, this comprises of 1206 total energy calculations ($2*3N_{atoms}$). The second order derivatives of the total energy with respect to the position of the ions are then calculated using central difference method, which is used to construct the force constant matrix, which then generates the dynamical matrix. The eigenvalues of the dynamical matrix are the phonon frequencies. The results are compared with the experimental inelastic neutron scattering results from [30]. Although the calculated VDOS reproduces most of the peaks observed in experimental VDOS, the peaks are shifted by ~70 cm$^{-1}$ towards lower wavenumbers, which means that the bond-lengths are slightly larger in our PBE relaxed model. This also leads to slightly larger volume of the relaxed simulation cell and a smaller density which can be seen in Table 3. This discrepancy results from the exchange-correlation functional as has already been shown for the specific case of the infrared spectrum [55]. Gradient corrections have the general tendency to overcorrect bond lengths [56] with the clear effect that all vibrational modes and, specifically, the stretching ones at higher frequencies, become slightly "slower" thus giving rise to the observed redshift.

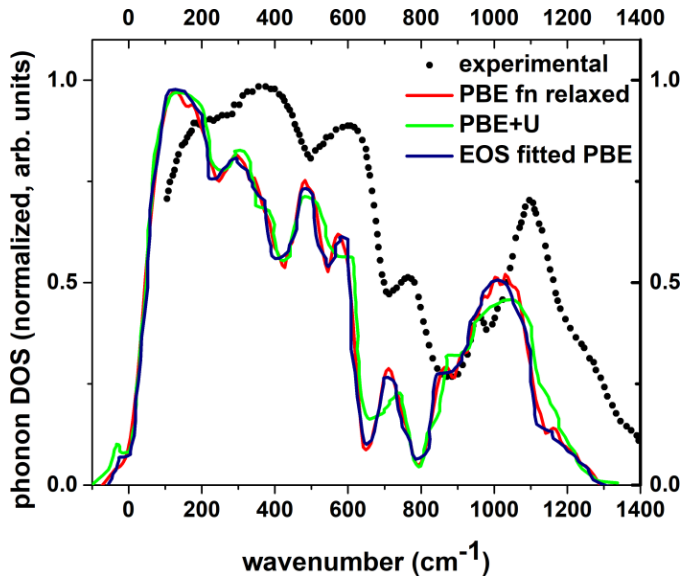

Figure 12: The vibrational density of states of IPG as obtained by various XC functionals and through inelastic neutron scattering experiments [30]

Partial phonon density of states: Also calculated was atom projected phonon density of states (shown in Figure 13) to understand the contribution of atoms in different regions of the frequency spectrum. The lightest of the atoms i.e., oxygen is found to contribute to all the frequencies found in the total DOS. The heaviest atom i.e. Fe is found to contribute only towards low frequency region (0-400 cm$^{-1}$) which is expected since the phonon frequency is inversely proportional to mass of the atom. The P atom, being lighter is also found to contribute at all the frequencies in total DOS. The frequencies in 700-1200 cm$^{-1}$ are contributed solely by phosphate group in the network.

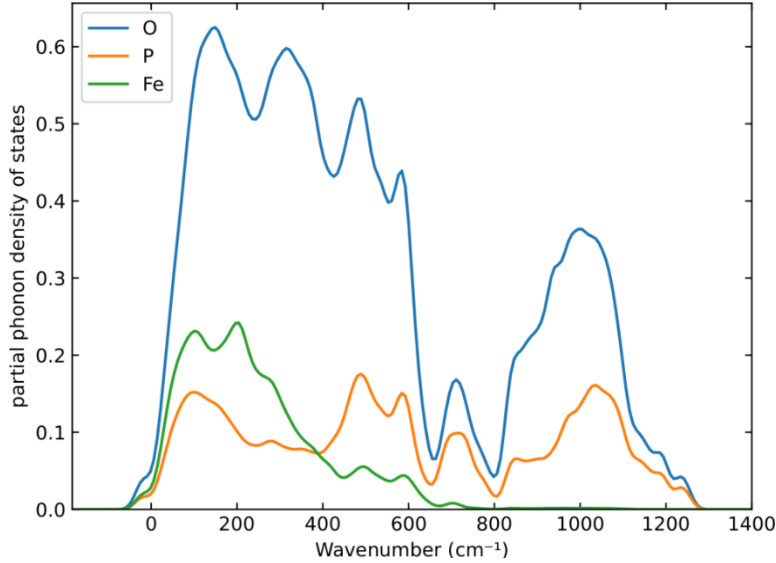

Figure 13: The atom-projected phonon density of states of IPG computed using PBE functional

### B. Elastic Constants

The calculation of second order force constants allows for the estimation of elastic constants which are presented next. These calculations require long computational time and hence have been restricted to only the PBE relaxed model. For comparison, the predicted values from classical MD (GULP [16] and LAMMPS (unpublished)) and experimental values [57] are also reported. Table 5 presents the coefficients of the stiffness matrix. As for any homogeneous and isotropic system, for the glassy system considered here, IPG, average values of the three independent coefficients i.e., $<C_{11}=C_{22}=C_{33}>$, $<C_{12}=C_{13}=C_{23}>$ and $<C_{44}=C_{55}=C_{66}>$ are reported. Other coefficients are found to be quite small - close to zero - as should be the case for a glassy system. The full stiffness matrix is provided in the Supplementary text (section 3). The values of the coefficients obtained in this study are in closer agreement with experimental data [57] than classical MD results, which boosts the scope of ab-initio modeling for accurate determination of these properties.

Table 6 presents the mechanical properties i.e bulk modulus, Young's modulus, shear modulus, Poisson's ratio of IPG along with the densities of simulation box, which strongly affects the mechanical properties. The values are calculated using ELATE [58]. The reported values are calculated using Voigt averaging scheme [59]. The elastic tensor is found to be positive definite i.e. all the eigenvalues of the matrix are positive (reported in supplementary file), indicating a mechanically stable system [60]. Figure S2 and Figure S3 plots the spatial variation of Young's modulus and linear compressibility. Comparison has been done with available experimental data and available theoretical data from MD simulations. The agreement of the DFT calculated values with experimental data is remarkable. The MD results show high dependence on the value of density chosen for simulation as well as the ensemble chosen while quenching and equilibration. From our parallel research on IPG using classical MD (unpublished), it was found that high quench rates, optimal density and choice of ensemble while simulating the melt-quench are still outstanding issues in MD simulation of IPG.

Table 5: Comparison of coefficients of Stiffness matrix, $C_{ij}$ of IPG from this study with experimental data

| Calculation method | $<C_{11} = C_{22} = C_{33}>$ (GPa) | $<C_{12} = C_{13} = C_{23}>$ (GPa) | $<C_{44} = C_{55} = C_{66}>$ (GPa) |
|---|---|---|---|
| DFT-VASP (this study) | 76.45 | 30.33 | 31.63 |
| MD-LAMMPS (unpublished) | 110.9 | 61.70 | 24.60 |
| Exptl. [57] | 85.90 | 27.80 | 29.0 |

Table 6: Comparison of mechanical properties for IPG obtained in this study with experimental data and other theoretical studies

| Calculation method | Density ρ (g/cc) | Bulk mod. B (GPa) | Young's mod. E (GPa) | Shear mod. S (GPa) | Poisson's ratio σ |
|---|---|---|---|---|---|
| DFT-VASP (this study) | 2.80 (PBE) | 45.7 | 70.2 | 28.2 | 0.24 |
| MD-LAMMPS (unpublished) | 3.12 | 78.10 | 68.23 | 25.27 | 0.35 |
| MD-GULP [15] | 3.25 | 75 | 106 | - | - |
| MD-GULP Lattice Dynamics [16] | 2.90 | 48.0 | 81.0 | 31.0 | 0.15 |

| Exptl. | 2.90 [61], 2.91 [57], 3.06 [52] | 47.2 [57] | 69.5 [10], 72.3 [57] | 29.0 (calc.) | 0.24 [57] |

## 4. Optical properties

The optical properties of IPG are presented next. Calculations are done in the independent particle approximation on model relaxed using hybrid density functional with U correction [62]. The linear optical properties of a system can be obtained from the frequency-dependent complex dielectric function $\varepsilon(\omega)$. We used VASPKIT [63] for post processing the output files. The relations governing these calculations are given in [63]. One important observation is that all the optical properties are isotropic (similar trend along *xx, yy* and *zz*) which is how it should be for a random-network glass. The calculated complex dielectric constant of IPG is plotted in supplementary file Figure S4. The value of the static dielectric constant is $\varepsilon_0 \approx 2.4$ and that of the high frequency dielectric constant is $\varepsilon_\infty \approx 1$.

The frequency-dependent linear optical spectra are shown in Figure 14. In (a) the absorption coefficient of IPG is shown, with very small absorption in the visible light range (1.7 to 3.3 eV). For the sake of completion, graphs for reflectivity $R(\omega)$, refractive index $n(\omega)$ and extinction coefficient $\kappa(\omega)$ are also shown in Figure 14 (b), (c) and (d), respectively.

For amorphous semiconductors/glasses such as IPG, band-gap is obtained by extrapolating a linear fit in the $(\alpha.h\nu)^n$ vs photon energy $(h\nu)$ plot. The value of exponent n of $\alpha.h\nu$ is ½ for the indirect optical band-gap and 2 for the direct optical band-gap corresponding to indirect and direct allowed transitions. Figure 15 (a), (b) and (c) show the Tauc's plot for absorption coefficient, $\alpha$ along xx, yy and zz respectively. The obtained indirect $E_g$ values are reported in the corresponding plots giving an average $E_g$ of 3.78 eV. The value is higher than the experimentally observed value of approx. 3.0 eV. The reason for this is because the optical properties are calculated using the U correction and hybrid HSE06 functional. The combined effect of this overestimates band-gap. The value of direct band-gap obtained from similar treatment in $(\alpha.h\nu)^2$ vs $h\nu$ plot comes out to be ~5.1 eV (shown in supplementary Figure S5).

Also calculated in Figure 15 (d) is the Urbach energy $E_u$ of this IPG model. Inverse of the slope of *ln α* vs *hν* graph gives the Urbach energy of a system, which gives a measure of the disorder in the system. The disorder may have thermal, structural or compositional origin. Equations pertaining to calculation of $E_u$ and its relation to structural disorder is given in supplementary file. Due to structural disorder in glasses, the EDOS near band-edges have an exponential decay as visible in Figure 9. This leads to exponential tail in absorption spectrum as well. The contribution to this band tailing effect is not only from the absence of long-range order but also from the presence of coordination defect states lying in the tail and band-gap region [45,46] (see supplementary file for relating $E_u$ with disorder). From Figure 9, the presence of gap-states in EDOS can be observed to be quite high, which lead to localized states with high IPR near the band-gap and band-edges. The value of $E_u$ for the IPG model is 0.288 eV. While crystals have much smaller $E_u$ (of the order of meV), typical glassy systems have similar order as IPG. Only one experimental study [50] provides the value of Urbach energy for a nearby concentration of IPG, in which $E_u$ is found to vary from 0.17 to 0.22 eV as concentration of $Fe^{2+}$ is varied from 20% to 0%. In our study, even when we assume the $Fe^{2+}$ concentration to be zero in our simulation cell and calculate $E_u$ at 0 K temperature thus removing any thermal contributions to the disorder, our value is still higher than the experimental one. This may be due to (1) the composition in this study is different, more iron oxide in the glass composition leads to higher structural disorder and (2) the hybrid density functional with U correction may also lead to over estimation of $E_u$. A more proper $Fe^{2+}$ concentration dependent study has to be carried out in the future.

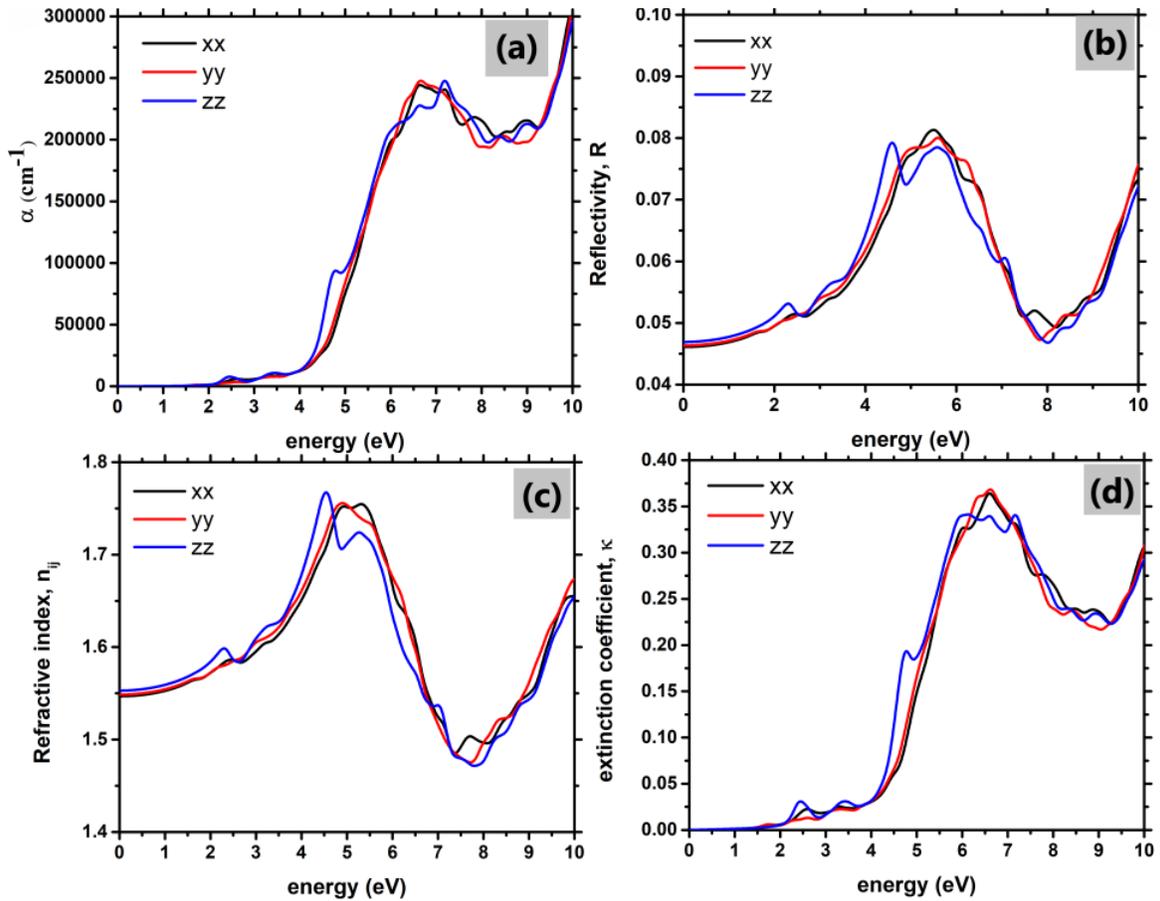

Figure 14: (a) Absorption coefficient (α), (b) reflectivity (R), (c) refractive index (n) and (d) extinction coefficient (κ) of IPG as a function of energy of incident light

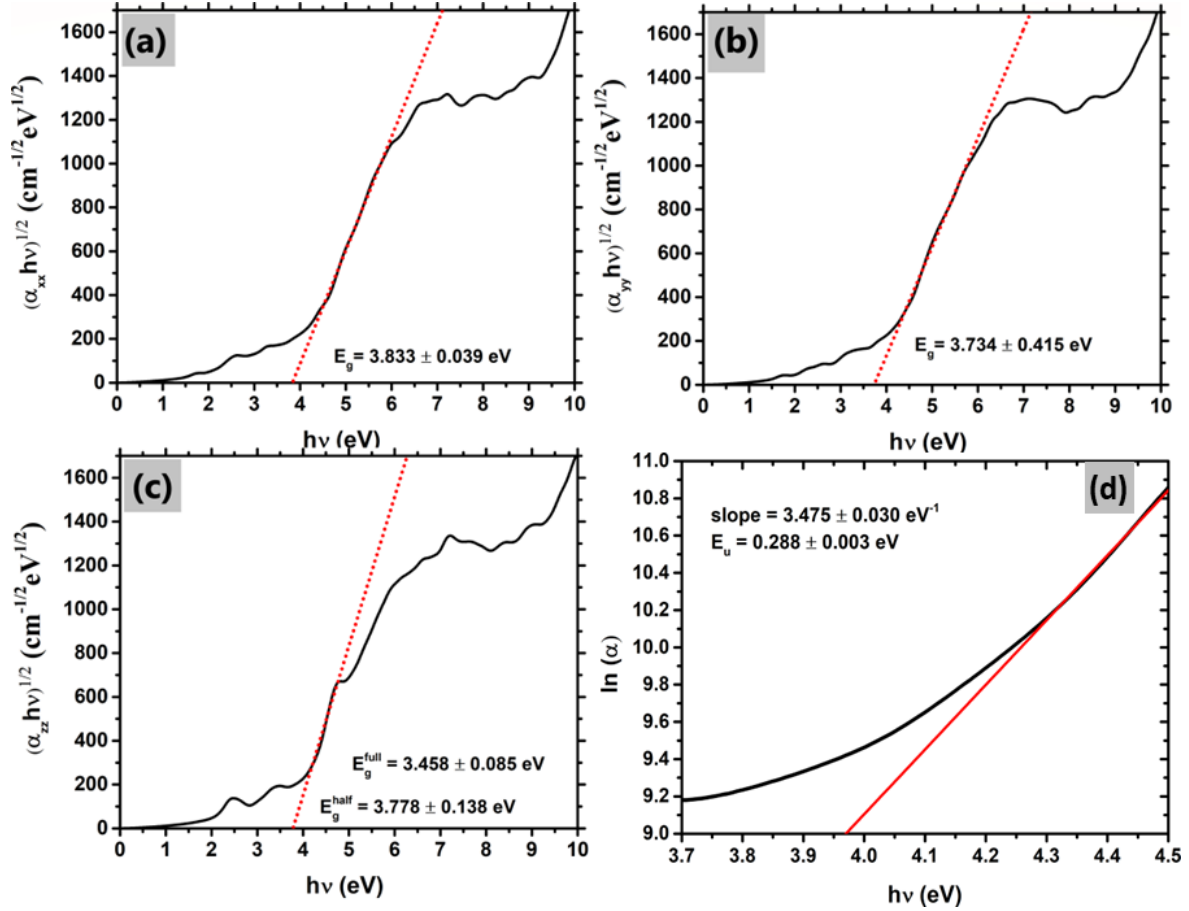

Figure 15: Tauc's plot for HSE + U relaxed model of IPG, using (a)$\alpha_{xx}$, (b) $\alpha_{yy}$, (c) $\alpha_{zz}$ and estimation of the Urbach energy $E_u$ from ln ($\alpha_{zz}$) in (d)

## IV. Conclusions

For finding the electronic structure of a glassy system, one must start the calculations on a plausible model, the technique to find such a model is an open problem which is being addressed for the case of IPG in this paper. We have obtained realistic models of IPG which are based on structural characterization done in this study. These are shown to reproduce the experimentally observed values of not only short-range properties like T(r), coordination number and bond-angle distribution but also intermediate range properties like FSDP in structure factor and void distribution. After structurally validating the models, the electronic properties of IPG such as EDOS, band-gap, IPR, magnetization, mechanical properties such as elastic constants and VDOS, and optical properties such as linear optical response and Urbach energy have been delineated. The successful agreement with experimental data for properties such as FSDP, band-gap, magnetization, VDOS, elastic constants and Urbach-edge have been reported for the first time for an ab-initio study on IPG. The hybrid density functional is found to reproduce the electronic structure of IPG most accurately as concluded from the DOS plots. The PBE functional calculations yields slightly larger bond-lengths and hence simulation cell volume leading to a slightly red-shifted vibrational DOS plot. The use of HSE functional together with U correction, gives slightly over-estimated band-gap values which is expected. The high Urbach energy of IPG is explained in terms of the structural disorder and defects present in glasses whose contribution to band-gap region is also established through EDOS and IPR plots. A satisfactory match with experimental data is reported throughout the study wherever available. The presented study also validates the hybrid method of generating IPG glass models employed in this study. A shortcoming was also reported in the form of incorrect ring distribution predominantly due to the small size of these models. Furthermore, the customizability of the MC method employed in this study, allows us to model compositionally different structures, and explore composition-property relation for IPG, which can be explored in future.

**Supplementary material**

The supplementary material contains various relations for calculations of radial distribution function, coordination geometries in IPG model, elastic properties of IPG, equations for calculating the frequency-dependent linear optical response, plot of real and imaginary components of dielectric function, calculation of direct optical band-gap, Urbach energy and its relation with structural disorder and magnetization on various atoms in the presented models. Supplementary data associated with this article can be found in the online version.

**Acknowledgements**

S.S thanks Dr. R Rajaraman, MSG, IGCAR and Dr. Kitheri Joseph of MC&MFCG, IGCAR for useful suggestions, S.S immensely thanks Dr. Gurpreet Kaur of MSG, IGCAR for DFT guidance, S.S, M.D and S.C thank the computer division, IGCAR for easy accessibility of the computational facilities.

**Author contribution**

S.S—computational work, data analysis, interpretation, writing, M.D—guidance, discussions and editing. S.C—problem definition, guidance, discussions and editing.

**Data availability statement**

The raw/processed data required to reproduce these findings cannot be shared at this time as the data also forms part of an ongoing study.

**Highlights/Synopsis:**

- Models of Iron phosphate glass (IPG) obtained using ab-initio method i.e. DFT
- Hybrid method comprising random-network modelling by Monte-Carlo method followed by ab-initio MD annealing then followed by DFT geometry optimization
- The models after structural validation, used for electronic structure calculation employing spin-polarised PBE and HSE functionals with and w/o U correction for Fe.
- EDOS, band-gap, VDOS, magnetism, elastic-constants, optical properties are reported for IPG

**For Table of Content Graphics only:**

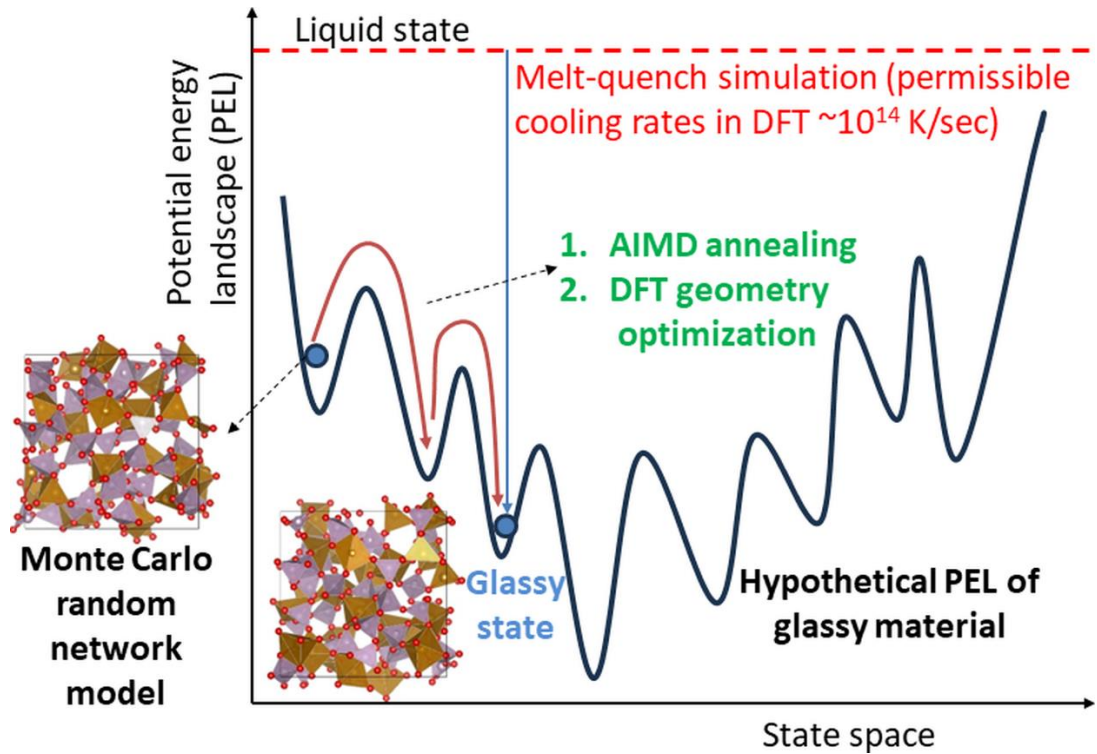

# Towards atomistic understanding of Iron phosphate glass: a first-principle based DFT modeling and study of its physical properties

Shakti Singh[a,*,‡], Manan Dholakia[a] and Sharat Chandra[a]

[a]Materials Science Group, Indira Gandhi Centre for Atomic Research, a CI of Homi Bhabha National Institute (Mumbai), Kalpakkam, TN, 603102 India

[‡]Present Address: Laser Biomedical Applications Division, Raja Ramanna Centre for Advanced Technology, a CI of Homi Bhabha National Institute (Mumbai), Indore, MP, 452013, India

E-mail: [*]corresponding author S.S: shaktisinghstephen@gmail.com,

M.D: manan@igcar.gov.on, S.C: sharat.c@gmail.com

## SUPPLIMENTARY INFORMATION

1. **Various definitions of radial distribution used in this study as reported in R.I.N.G.S. software [1] manual [https://sourceforge.net/projects/rings-code]:**

A. Partial radial distribution function, $g_{\alpha\beta}(r)$:

$$g_{\alpha\beta}(r) = \frac{dn_{\alpha\beta}(r)}{4\pi r^2 dr\, \rho_\alpha}$$

where $\rho_\alpha = \frac{V}{N_\alpha} = \frac{V}{N*c_\alpha}$ with V, the volume of simulation cell, N is the total number of atoms, $N_\alpha$ is the number of atoms of specie $\alpha$, $c_\alpha = \frac{N_\alpha}{N}$, is the concentration of atomic species $\alpha$.

B. Radial distribution function also called pair correlation function, $g(r)$:

$$g(r) = \sum_{\alpha,\beta} \frac{c_\alpha b_\alpha c_\beta b_\beta g_{\alpha\beta}(r)}{\langle b^2 \rangle}$$

where $b_\alpha$ is the neutron or X-ray scattering length of species $\alpha$, $\langle b^2 \rangle = (\sum_\alpha c_\alpha b_\alpha)^2$ and $g_{\alpha\beta}(r)$ as calculated from (A).

C. Reduced pair correlation function, $G(r)$:

$$G(r) = \sum_{\alpha,\beta} c_\alpha b_\alpha c_\beta b_\beta (g_{\alpha\beta}(r) - 1)$$

where $g_{\alpha\beta}(r)$ can be calculated from (A).

D. Total pair correlation function, $T(r)$:

$$T(r) = 4\pi r \rho \, (G(r) + \langle b^2 \rangle)$$

where $G(r)$ can be obtained from (C).

E. Neutron scattering static structure factor, $S(q)$

$$S(q) = \sum_\alpha c_\alpha b_\alpha^2 + I(q)$$

where $I(q)$ describes the component of structure factor from interaction between distinct atoms, $q$ is the scattering wave vector, $b_\alpha$ is the neutron scattering length of species $\alpha$. $I(q)$ is directly obtainable from Fourier transform of reduced pair correlation function, $G(r)$ defined in (C) as follows:

$$I(q) = 4\pi\rho \int_0^\infty dr \, r^2 \, \frac{\sin qr}{qr} G(r)$$

2. **Coordination geometries in IPG model:**

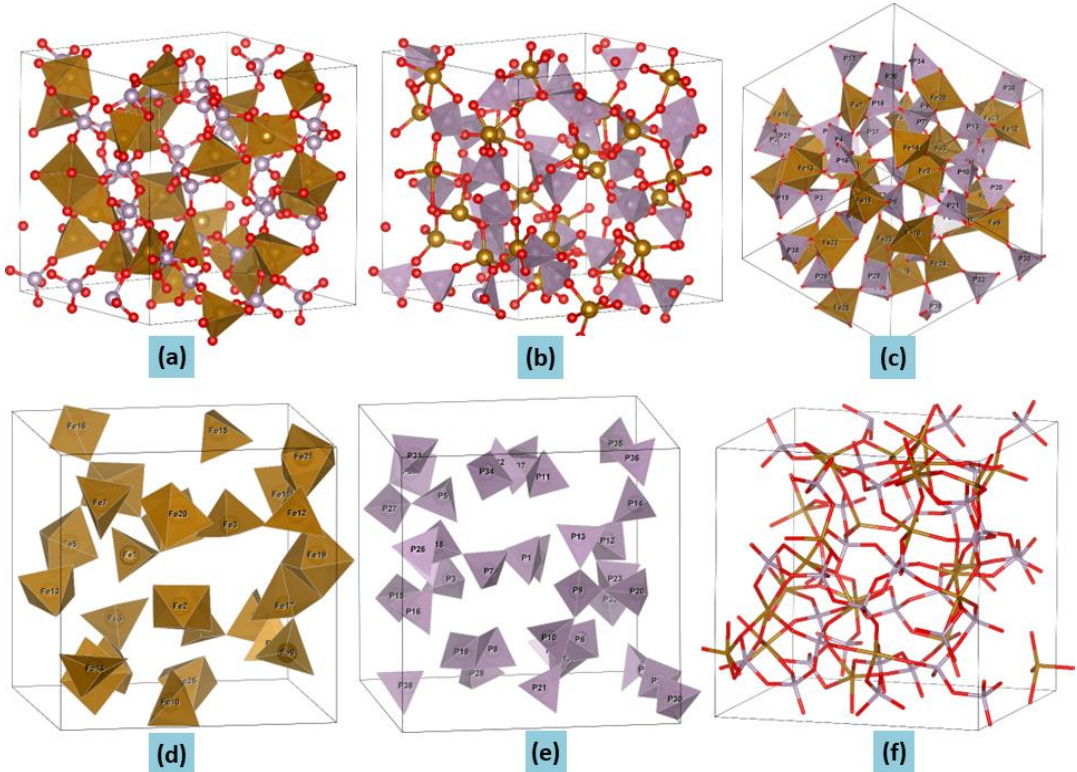

Figure S1: Snapshots of IPG model obtained from HSE06 functional relaxation: (a) shows only Fe polyhedral units (b) only P tetrahedral in the network, (c) shows all coordination geometries, (d) and (e) shows the coordination geometries of Fe and P respectively without oxygen and its network, (f) shows the continuous random-network of IPG using only bond representation (red: O , violet: P, golden: Fe contribution)

3. **Elastic properties of IPG:**

- Stiffness matrix (coefficients in GPa) of IPG

$$[C_{ij}] = \begin{bmatrix} 95.17 & 28.77 & 22.31 & -3.27 & 6.33 & 0.03 \\ 28.77 & 61.10 & 39.90 & 5.42 & 9.69 & -2.14 \\ 22.31 & 39.90 & 73.09 & -10.99 & 2.60 & 1.78 \\ -3.27 & 5.42 & -10.99 & 31.92 & 1.72 & 0.17 \\ 6.33 & 9.69 & 2.60 & 1.72 & 28.88 & 1.95 \\ 0.03 & -2.14 & 1.78 & 0.17 & 1.95 & 34.08 \end{bmatrix}$$

- Eigenvalues of the stiffness matrix
  **(calculated from  https://progs.coudert.name/elate [2])**

| $\lambda_1$ | $\lambda_2$ | $\lambda_3$ | $\lambda_4$ | $\lambda_5$ | $\lambda_6$ |
|---|---|---|---|---|---|
| 15.832 | 26.468 | 34.758 | 42.15 | 66.048 | 138.99 |

- Spatial dependence of Young's modulus:

**(calculated from https://progs.coudert.name/elate [2])**

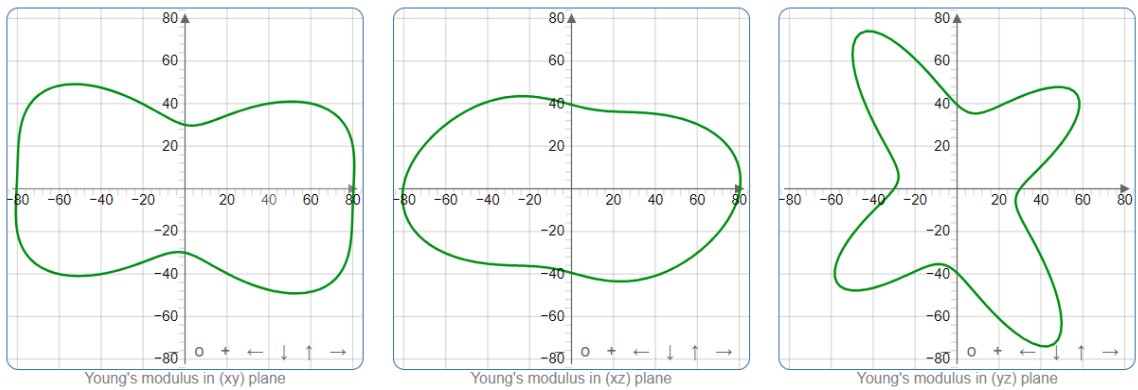

Figure S2: (From left) Variation of Young's modulus in xy, xz and yz plane respectively in the glass

- Spatial dependence of linear compressibility
  **(calculated from https://progs.coudert.name/elate [2])**

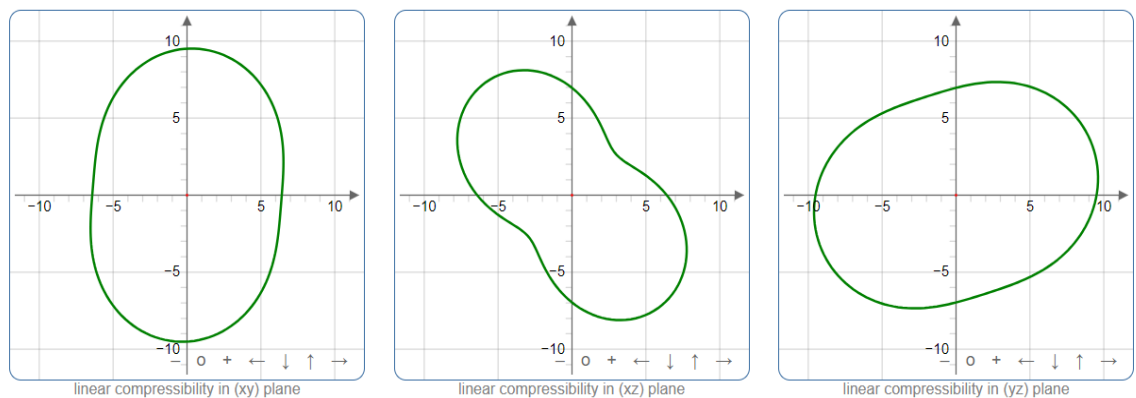

Figure S3: (From left) Variation of linear compressibility in xy, xz and yz plane respectively in the glass

4. **Equations for calculating the frequency-dependent linear optical response.**

- Dielectric function, $\varepsilon$

$$\varepsilon(\omega) = \varepsilon_1(\omega) + i\, \varepsilon_2(\omega)$$

where $\varepsilon_1(\omega)$ and $\varepsilon_2(\omega)$ are the real and imaginary parts of the dielectric function, and $\omega$ is the photon frequency.

- Absorption coefficient, $\alpha$

$$\alpha(\omega) = \sqrt{\frac{2\omega}{c}}\left((\varepsilon_1^2 + \varepsilon_2^2)^{1/2} - \varepsilon_1\right)^{1/2}$$

- Refractive index, n

$$n(\omega) = \left(\frac{(\varepsilon_1^2 + \varepsilon_2^2)^{\frac{1}{2}} + \varepsilon_1}{2}\right)^{1/2}$$

- Extinction coefficient, κ

$$\kappa(\omega) = \left(\frac{(\varepsilon_1^2 + \varepsilon_2^2)^{\frac{1}{2}} - \varepsilon_1}{2}\right)^{1/2}$$

- Reflectivity, R

$$R(\omega) = \left(\frac{(n-1)^2 + \kappa^2}{(n+1)^2 + \kappa^2}\right)$$

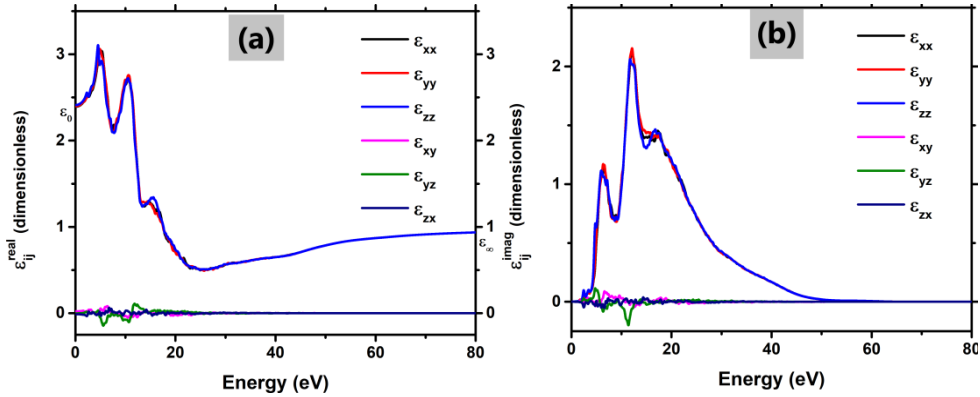

Figure S4: The calculated dielectric function of IPG (a) real and (b) imaginary part

- **Calculation of direct optical band-gap of IPG**

Figure S5 shows the plot of $(\alpha_{xx}E)^2 \ vs \ E$, for model relaxed using HSE+U. The direct band-gap obtained from the extrapolation of the linear region gives a value of 5.09 eV.

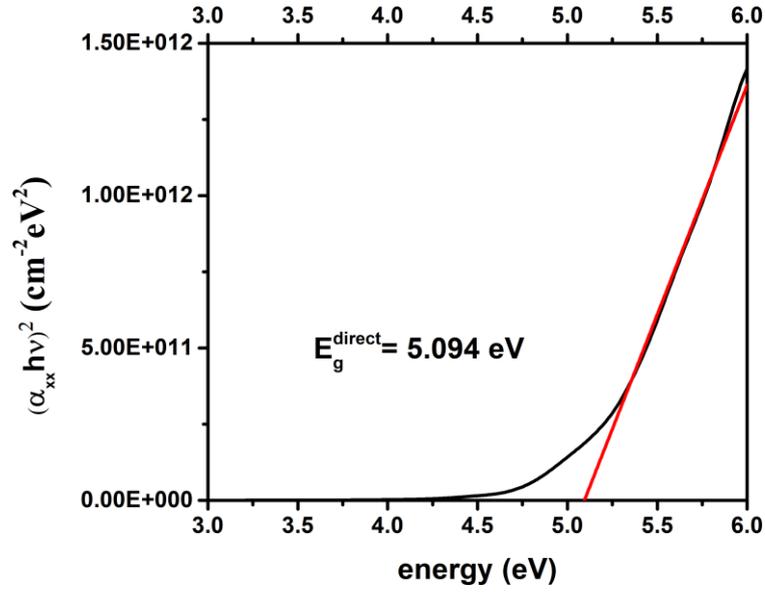

Figure S5: Direct optical band-gap of IPG from absorption plot

- **Calculation of Urbach energy and relation with structural disorder**

Absorption in semiconductors is known to increase exponentially near the onset of absorption, spanning several orders of magnitude. Absorption as a function of energy can be described by the following equation (famously the Urbach rule) [3,4]:

$$\alpha(E) = \alpha_0 * exp\left(\frac{E - E_g}{E_u}\right)$$

where $\alpha_0$ and $E_g$ are fitting parameters with dimensions of inverse length and energy, respectively, and $E_u$ is the Urbach Energy. This equation is only valid when alpha is proportional to exp(E) i.e. in the band-tail. Taking log on both sides of above equation

$$ln\,\alpha = ln\,\alpha_0 + \left(\frac{E - E_g}{E_u}\right)$$

$$ln\,\alpha = constant + \left(\frac{E}{E_u}\right)$$

So, the slope of $ln\,\alpha$ vs E graph near the band-gap region will give the inverse of Urbach energy, $E_u$.

- **Inference about the structural disorder from $E_u$:**

Contribution to $E_u$ can be written as [4]:

$$E_u(T) = K_B(\langle U^2 \rangle_{thermal} + \langle U^2 \rangle_{structural})$$

where $\langle U^2 \rangle_{structural}$ is the contribution of structural (topological) disorder to the mean-square deviation of the atomic positions from a perfectly ordered configuration. Similar contribution from thermal vibrations is $\langle U^2 \rangle_{thermal}$. At 0 K, where the calculations of optical properties is done in VASP, we only have one term (for detailed explanation see [4]):

$$E_u(0) = \frac{K_B \theta}{\sigma_0} \left( \frac{1 + X}{2} \right)$$

where $\sigma_0$ is the Urbach-edge parameter of order unity, $X = \frac{\langle U^2 \rangle_{structural}}{\langle U^2 \rangle_0}$, where $X$ is the measure of structural disorder normalised to $\langle U^2 \rangle_0$, the zero-point uncertainty in atomic positions and $\theta$ is the characteristic temperature of the Einstein oscillator used for approximating the phonon spectrum of the material. In the absence of experimental values of parameters like $\sigma_0$, $\theta$ which are not available from literature for this system, only inference about the proportionality of the $E_u$ to $X$ can be made i.e. **more the Urbach energy more will be the structural disorder**.

5. **Magnetisation:**

Table S1 and Table S2 reports the DFT calculated values of magnetisation along x axis for different atoms in the proposed models.

Table S1: The calculated magnetisation on atoms in IPG. For atoms O and P, the average magnetisation over all O, <O> and over all P, <P> is reported. For Fe, magnetisation on all atoms and average over all Fe, <Fe> is also reported. This table reports values for PBE and PBE+U relaxed models

| Atom type | Magnetisation(x) | | | | | | | |
|---|---|---|---|---|---|---|---|---|
| | PBE relaxed model | | | | PBE+U relaxed model | | | |
| | s | p | d | tot | s | p | d | tot |
| <O> | 0.01 | 0.09 | 0.00 | 0.10 | 0.01 | 0.08 | 0.00 | 0.08 |
| <P> | 0.01 | 0.01 | 0.01 | 0.03 | 0.00 | 0.01 | 0.01 | 0.03 |
| Fe | 0.03 | 0.06 | 3.72 | 3.81 | 0.03 | 0.06 | 4.29 | 4.38 |
| Fe | 0.02 | 0.03 | 3.99 | 4.04 | 0.02 | 0.03 | 4.45 | 4.50 |
| Fe | 0.02 | 0.04 | 3.93 | 3.99 | 0.02 | 0.04 | 4.36 | 4.43 |
| Fe | 0.03 | 0.03 | 2.54 | 2.61 | 0.03 | 0.05 | 4.31 | 4.39 |
| Fe | 0.02 | 0.02 | 3.94 | 3.98 | 0.02 | 0.03 | 4.37 | 4.42 |
| Fe | -0.01 | -0.03 | -3.73 | -3.78 | 0.02 | 0.04 | 4.25 | 4.31 |
| Fe | 0.02 | 0.04 | 3.92 | 3.98 | 0.02 | 0.04 | 4.36 | 4.42 |
| Fe | 0.03 | 0.04 | 2.59 | 2.66 | 0.03 | 0.05 | 4.15 | 4.23 |
| Fe | 0.02 | 0.07 | 3.81 | 3.90 | 0.03 | 0.07 | 4.28 | 4.37 |
| Fe | 0.02 | 0.04 | 3.93 | 3.99 | 0.02 | 0.04 | 4.35 | 4.42 |
| Fe | 0.03 | 0.06 | 3.88 | 3.96 | 0.03 | 0.06 | 4.25 | 4.33 |
| Fe | 0.02 | 0.07 | 3.85 | 3.93 | 0.02 | 0.06 | 4.29 | 4.38 |
| Fe | 0.02 | 0.05 | 3.89 | 3.97 | 0.03 | 0.05 | 4.29 | 4.37 |
| Fe | 0.03 | 0.04 | 3.70 | 3.77 | -0.03 | -0.04 | -4.20 | -4.27 |
| Fe | 0.02 | 0.07 | 3.83 | 3.92 | 0.02 | 0.07 | 4.30 | 4.39 |
| Fe | 0.03 | 0.03 | 2.89 | 2.96 | 0.03 | 0.05 | 3.85 | 3.93 |
| Fe | 0.05 | 0.03 | 3.68 | 3.75 | 0.04 | 0.02 | 3.81 | 3.87 |

| Fe | 0.02 | 0.05 | 3.89 | 3.97 | 0.03 | 0.05 | 4.35 | 4.42 |
| Fe | 0.02 | 0.04 | 3.94 | 3.99 | -0.02 | -0.03 | -4.36 | -4.41 |
| Fe | -0.02 | -0.03 | -3.30 | -3.36 | 0.03 | 0.04 | 4.22 | 4.28 |
| Fe | -0.01 | -0.01 | -2.00 | -2.02 | 0.02 | 0.03 | 4.36 | 4.42 |
| Fe | 0.02 | 0.04 | 3.83 | 3.89 | 0.03 | 0.05 | 4.17 | 4.25 |
| Fe | 0.02 | 0.06 | 3.19 | 3.27 | 0.03 | 0.06 | 4.25 | 4.34 |
| Fe | 0.02 | 0.05 | 3.60 | 3.68 | 0.02 | 0.07 | 4.27 | 4.36 |
| Fe | 0.02 | 0.05 | 3.92 | 3.98 | 0.02 | 0.05 | 4.39 | 4.46 |
| Fe | 0.03 | 0.03 | 2.58 | 2.64 | 0.03 | 0.03 | 4.26 | 4.33 |
| **<Fe>** | **0.02** | **0.04** | **2.85** | **2.90** | **0.02** | **0.04** | **3.60** | **3.67** |

Table S2: The calculated magnetisation on atoms in IPG. For atoms O and P, the average magnetisation over all O, <O> and over all P, <P> is reported. For Fe, magnetisation on all atoms and average over all Fe, <Fe> is also reported. This table reports values for HSE and HSE+U relaxed models

| Atom type | Magnetization(x) | | | | | | | |
|---|---|---|---|---|---|---|---|---|
| | HSE relaxed model | | | | HSE+U relaxed model | | | |
| | s | p | d | tot | s | p | d | tot |
| <O> | 0.01 | 0.09 | 0.00 | 0.10 | 0.01 | 0.06 | 0.00 | 0.07 |
| <P> | 0.01 | 0.01 | 0.01 | 0.03 | 0.00 | 0.01 | 0.01 | 0.03 |
| Fe | 0.02 | 0.05 | 4.07 | 4.15 | 0.03 | 0.05 | 4.51 | 4.59 |
| Fe | 0.01 | 0.02 | 4.23 | 4.27 | 0.02 | 0.02 | 4.64 | 4.68 |
| Fe | 0.01 | 0.04 | 4.15 | 4.20 | 0.02 | 0.04 | 4.57 | 4.63 |
| Fe | 0.03 | 0.05 | 4.06 | 4.14 | 0.03 | 0.05 | 4.51 | 4.59 |
| Fe | 0.02 | 0.02 | 4.21 | 4.25 | 0.02 | 0.02 | 4.58 | 4.63 |
| Fe | 0.02 | 0.04 | 4.12 | 4.19 | 0.03 | 0.05 | 4.52 | 4.59 |
| Fe | 0.01 | 0.04 | 4.15 | 4.20 | 0.02 | 0.04 | 4.56 | 4.62 |
| Fe | 0.03 | 0.05 | 4.05 | 4.12 | 0.03 | 0.05 | 4.45 | 4.53 |
| Fe | 0.02 | 0.07 | 4.04 | 4.13 | 0.02 | 0.07 | 4.49 | 4.58 |
| Fe | 0.02 | 0.03 | 4.17 | 4.22 | 0.02 | 0.03 | 4.58 | 4.63 |
| Fe | 0.02 | 0.06 | 4.04 | 4.11 | 0.02 | 0.06 | 4.45 | 4.53 |
| Fe | 0.01 | 0.07 | 4.06 | 4.14 | 0.02 | 0.07 | 4.50 | 4.59 |
| Fe | 0.02 | 0.05 | 4.08 | 4.15 | 0.02 | 0.05 | 4.51 | 4.58 |
| Fe | 0.04 | 0.05 | 4.06 | 4.15 | 0.04 | 0.05 | 4.46 | 4.56 |
| Fe | 0.01 | 0.07 | 4.06 | 4.14 | 0.02 | 0.06 | 4.50 | 4.58 |
| Fe | 0.03 | 0.05 | 3.43 | 3.50 | 0.03 | 0.04 | 4.48 | 4.55 |
| Fe | 0.03 | 0.02 | 3.62 | 3.66 | 0.03 | 0.02 | 3.84 | 3.89 |
| Fe | 0.02 | 0.04 | 4.12 | 4.18 | 0.02 | 0.05 | 4.54 | 4.61 |
| Fe | 0.01 | 0.03 | 4.17 | 4.21 | 0.02 | 0.03 | 4.58 | 4.63 |
| Fe | 0.02 | 0.04 | 4.03 | 4.09 | 0.03 | 0.05 | 4.50 | 4.58 |
| Fe | 0.01 | 0.03 | 4.17 | 4.21 | 0.02 | 0.03 | 4.57 | 4.62 |
| Fe | 0.01 | 0.04 | 4.07 | 4.12 | 0.02 | 0.04 | 4.49 | 4.56 |
| Fe | 0.02 | 0.06 | 4.03 | 4.11 | 0.03 | 0.06 | 4.46 | 4.55 |
| Fe | 0.01 | 0.07 | 4.04 | 4.12 | 0.02 | 0.07 | 4.49 | 4.57 |
| Fe | 0.01 | 0.04 | 4.15 | 4.21 | 0.02 | 0.04 | 4.59 | 4.66 |
| Fe | 0.03 | 0.02 | 4.08 | 4.14 | 0.03 | 0.03 | 4.48 | 4.54 |
| **<Fe>** | **0.02** | **0.04** | **4.06** | **4.12** | **0.02** | **0.04** | **4.49** | **4.56** |